\documentclass[preprint]{revtex4-1}

\usepackage{graphicx}
\usepackage{dcolumn}
\usepackage{bm}
\usepackage{natbib}

\usepackage{setspace}
\usepackage{epsf,epsfig,epstopdf,graphicx,amssymb,fancyhdr,url}
\usepackage{amssymb, amsmath, amsthm}
\usepackage{multirow}

\newcommand{\1}{\left}

\begin{document}

\title{Potentiation Decay of Synapses and the Length Distributions of Synfire Chains Self-organized in Recurrent Neural Networks}
\author{Aaron Miller}
\email{ajmiller@bridgewater.edu}
\affiliation{Department of Physics, Bridgewater College, Bridgewater, Virginia 22812, USA}
\author{Dezhe Z. Jin}
\email{djin@phys.psu.edu}
\affiliation{Department of Physics, The Pennsylvania State University, University Park, Pennsylvania 16802, USA}
\date{\today}

\begin{abstract}

Synfire chains are thought to underlie
precisely-timed sequences of spikes observed in various brain regions and across species. 
How they are formed is not understood. 
Here we analyze self-organization of synfire chains through the spike-timing dependent plasticity (STDP) 
of the synapses, axon remodeling, and 
potentiation decay of synaptic weights 
in networks of neurons driven by 
noisy external inputs and subject to dominant feedback inhibition. 
Potentiation decay is the gradual, activity-independent reduction of synaptic weights over time. 
We show that potentiation decay enables a dynamic and statistically stable network connectivity when neurons spike spontaneously. 
Periodic stimulation of a subset of neurons leads to formation of synfire chains through a random 
recruitment process, which terminates when the chain connects to itself and forms a loop.  
We demonstrate that chain length distributions depend on the potentiation decay. 
Fast potentiation decay leads to long chains with wide distributions, 
while slow potentiation decay leads to short chains with narrow distributions. 
We suggest that the potentiation decay, which corresponds to the decay of early long-term potentiation of synapses (E-LTP), is an important synaptic plasticity rule in regulating formation 
of neural circuity through STDP. 
\end{abstract}

\pacs{87.18.Sn, 84.35.+i }


\maketitle

 \section{Introduction}

Coevolution of the dynamics and topology of networks is widely observed in diverse systems from 
cellular biology to social networks \cite{luscombe2004genomic, powell2005network, gross2008adaptive}.
In the brain, 
the spiking dynamics of neurons depends on how they are connected. 
On the other hand, the connectivity can be modified by the spiking activity. 
The connections  (synapses)  between neurons in many brain areas are modified according to the spike-timing-dependent 
plasticity (STDP) rule \cite{gerstner1996neuronal,markram1997regulation,bi1998synaptic,caporale2008spike}.
A most common STDP rule for excitatory neurons is as follows \cite{caporale2008spike}:
the connection strength (synaptic weight) from neurons A to B is strengthened if A spikes before B (long-term potentiation, or LTP), 
and weakened if A spikes after B (long-term depression, or LTD). 
The amount of modification decreases with the time difference between the spikes of A and B. 
The connection between two neurons is lost if the synaptic weight is reduced below a threshold; conversely, 
it can be established through consistent parings of the spikes of the neurons. 
Studies that use STDP in spiking neural networks have shown a number of emergent properties
\cite{levy2001distributed,kang2004self,shin2006self,izhikovich2004spike,morrison2007spike,li2009self}.
In this paper, we show that synfire chain connectivity  \cite{abeles1982local,abeles1991corticonics}, 
in which subsequent groups of neurons are connected into a feedforward network
that supports sequential spiking of the neurons,  
emerges through coevolution of the spiking activity and the connectivity across many presentations of a training stimulus to a subset of neurons (training neurons).   

Sequential spiking of neurons is observed in a number of brain areas 
\cite{prut1998spatiotemporal,luczak2007sequential,hahnloser2002ultra}.
Some of the strongest experimental evidence for synfire chains producing spike sequences is from zebra finch premotor nucleus HVC (proper name)  \cite{hahnloser2002ultra,long2008using,long2010support}. 
Projection neurons in HVC spike sequentially at precise times relative to the learned song \cite{hahnloser2002ultra}.
Consistent with the synfire chain dynamics, 
cooling HVC uniformly slows down the song \cite{long2008using}, 
and the sub-threshold membrane potentials of neurons rapidly depolarize 5-10 ms before they spike \cite{long2010support}. 
It is well-established that synfire chains robustly
produce spike sequences 
\cite{abeles1982local,abeles1991corticonics,diesmann1999stable,jin2007intrinsic}.
However, how neurons are wired into synfire chains is not well-understood. 

An intriguing possibility is that synfire chains self-organize through activity-driven synaptic plasticity.  
Earlier studies using STDP or similar Hebbian rules 
resulted in short chains with a few groups \cite{hertz1996learning, levy2001distributed, kitano2002self}. 
The most likely reason is that these rules are prone to producing unstable growth of connections \cite{jun2007development,morrison2007spike,kunkel2010limits}. 
Two recent studies introduced additional homeostatic synaptic rules to limit such instability, and showed that long synfire chains can form  \cite{jun2007development,fiete2010spike}. 
The key idea behind both studies is to restrict the connectivity of the network after a certain amount of growth has
occurred.
Fiete et al achieved this by limiting the total synaptic weight in and out of every neuron \cite{fiete2010spike}.
However, a study using large scale simulations and mean-field analysis 
suggested that the regulation of the total synaptic weights
does not prevent unstable network growth \cite{kunkel2010limits}. 
In \cite{jun2007development} (Jun-Jin model), 
we took a different approach, and imposed an axon remodeling rule that limits the number $N_{ss}$ of strong connections, defined as those with synaptic weights above a threshold, that one neuron can maintain.  
Reaching the limit leads to pruning of all weak connections from the neuron.  
There are two additional features of the Jun-Jin model.
First, the model includes
an gradual, activity-independent decay of synaptic weights, which we call  potentiation decay.
Second, an activation threshold switches synapses on or off
depending on the magnitude of the synaptic weights.
This rule allows the active connections between neurons to form or disappear as the synaptic weights are modified. 
Simulations of 1000 leaky integrate-and-fire neurons showed that 
synfire chains emerge from initially random active connections when 6 to 40 training neurons are intermittently activated by external inputs for many trials. 
The number of neurons in each group roughly equals to $N_{ss}$ (set to 10 in the simulations), and is not affected by the number of training neurons except for the first 2-3 groups \cite{jun2007development}. 

In this paper, we perform an in-depth analysis of the Jun-Jin model. 
We address unresolved fundamental questions such as what determines the lengths of the emergent chains and how the length distributions are influenced by the total number of the neurons (network size).
We establish that, when the network is randomly active, the synaptic plasticity rules in this model allow the network connectivity to fluctuate, but the synaptic weights remain in a statistically 
stationary distribution.   
This ensures that the chain formation does not depend on specific initial network connectivity. 
We demonstrate that in between training trials, 
sequential spikes can spontaneously emerge in the forming chain.
This noise-induced re-activation of the chain creates connections from neurons outside of the chain to those in the chain,  
and plays a critical role in determining the length distributions of the final chains. 
Most notably, there is an upper limit for the mean chain length as the network size becomes large. 
We  show that slow potentiation decay leads to short chains with narrow length distributions, 
while fast potentiation decay leads to long chains with a wide length distributions. 
We compare the results of network simulations to a lottery-type stochastic process in which neurons are selected iteratively to enter a chain, 
and the chain stops growing when a loop is formed, either by selecting neurons 
already in the chain or by selecting neurons connected to the chain.
The distribution of chain
lengths from the network simulations fits well with distributions generated by the lottery process.
The analysis of this simple growth model shows that 
the rate of potentiation decay influences the chain length distributions by controlling the emergence of the connections to the growing chain.   
These connections also lead to a finite limit for the mean chain length as the network size increases. 

\section{Methods}

\subsection{Simulations}

We simulate synfire chain formation in recurrent networks of excitatory neurons.
We model each neuron using a leaky integrate-and-fire (LIF) model.
The neurons interact via pulse conductances, and they receive dominant feedback inhibition.
They are also
driven by upstream regions that we do not simulate, but instead model as independent, fluctuating external inputs.
Synaptic weights between neurons are modified according to an STDP rule. 
The details of modeling can be found in the Appendix.

Our model choices were dictated by necessity to simulate 
the network dynamics quickly
\cite{reutimann2003event,jin2002,tsodyks1993pattern}.
A large number of training trials ($10^4$ to $10^5$) are required for synfire chain formation in our model, and many training sessions are needed to construct the chain length distributions 
for a range of model parameters. 
Therefore it was necessary that our simulation algorithm be efficient.
We modified a fast, event-driven algorithm \cite{reutimann2003event} 
that had been developed to generate activity of pulse-coupled neurons that are targeted by
a fluctuating external input. When the external input is modeled by Gaussian white noise (GWN),
one can numerically solve the Fokker-Planck equation \cite{capocelli1971diffusion}, store particular solutions in 
``lookup tables'' and sample them during the network simulation 
to generate spike times. The steps of the algorithm are detailed in the Appendix.
The computational advantage of using pulse-coupled neurons is 
that the response of the membrane potential is instantaneous and
can be calculated exactly. The time-evolution
of the membrane potential between spikes is calculated from the lookup tables.

By our measurements, this algorithm is up to 150 times faster than simulating
with 4th-order Runge-Kutta method \cite{hansel1998numerical}. Instead of scaling
with number of timesteps, the simulation time scales with number of spikes, which results in increased simulation speed.
Two differences distinguish our simulation algorithm from that reported in \cite{reutimann2003event}.
First, we simulate conductance-based neurons instead of current-based neurons. Second, instead of a
spike latency, we impose a time resolution on the arrival times of spikes, as suggested in \cite{faisal2007stochastic}.
Algorithmically, imposing a time resolution on the spike arrival times means that, 
instead of the neuron with the single earliest predicted time emitting a spike,
any neurons that spike within an interval 
$T_{res}=3 \mbox{ms}$ of the earliest predicted spike time effectively spike together. 
The arrival time of the spikes at
synapses is picked to be at the end of the resolution interval. This method has the same effect as a random latency, 
allowing neurons
to cooperate to excite common targets, but it requires no additional queuing of events, which can be computationally
intensive and slows simulation considerably \cite{mattia2000efficient}.

The population of neurons we simulate make excitatory connections to each other. 
However, we assume a population of interneurons targets the excitatory population, and these neurons
reliably spike immediately when the excitatory neurons spike.
All of the neurons in the excitatory population are inhibited at the end of the resolution interval.
Near-global inhibition is observed in neocortical circuits  \cite{fino2011dense} and in the songbird premotor area HVC \cite{mooney2005hvc}.  
We do not simulate interneuron activity in order to
conserve computational resources.
The inhibition is stronger when more neurons spike within the spike resolution window,
but we put a constraint on it based on the assumption that
there is a finite size to the interneuron population targeting the simulated neurons \cite{hendry1987numbers}. 
Details are left to the Appendix.

\subsection{Synaptic weight dynamics}

The synaptic weight  between each pair of simulated neurons is modified based on the STDP rule (details are in the Appendix). 
Three additional synaptic plasticity rules are implemented to 
deter unchecked synaptic strengthening that STDP alone can lead to 
\cite{jun2007development,morrison2007spike,kunkel2010limits}, 
and to ensure stable synaptic weight distribution when the network is in the state of spontaneous activity with no training stimulations. 

\emph{Activation threshold:} 
Silent synapses are those with no post synaptic AMPA receptors \cite{kerchner2009silent}. 
At physiological conditions, these synapses do not produce responses in the postsynaptic neuron, hence ``silent". 
LTP can activate silent synapses to become functional synapses \cite{kerchner2009silent}; conversely, 
LTD can silence active synapses \cite{xiao2004creation,zhou2004shrinkage}. 
Abundant especially during the development, 
silent synapses allow the possibility of sculpting wide variety of neural circuits through neural activity. 

We model silencing and  activation of synapses by thresholding the synaptic weights. 
If synaptic weight from neuron $i$ onto neuron $j$, $G_{i,j}$, grows larger than $\Theta_A$,
then it is active and evokes a response from its target; otherwise it is silent and behaves as if it has zero weight. For our simulations, we pick $\Theta_A=.02$. 
Regardless of whether a synapse is silent or active, it obeys the STDP rule.
Our results do not depend on synaptic depression acting on
silent synapses.

Active synaptic connectivity \emph{in vivo} is sparse with a single neuron connecting to less than 10\% of its neighbors \cite{braitenberg1998cortex}.
The activation threshold directly avoids densely connected network states by deactivating all sufficiently weak connections.
A synapse between any pair of neurons can be
activated, so activity can drive the development of any possible synaptic connectivity.
In other words, there is no  \emph{a priori} restriction on how neurons can be connected after training.  
Parameters are selected such that the connectivity remains within the sparseness bounds observed experimentally.
A common modeling approach for avoiding dense connectivity is to specify a sparse connectivity between neurons and allow only synaptic weights of these connections to change \cite{izhikovich2004spike, morrison2007spike}. 
In this strategy no new connections can form, and the effects of training on the connectivity is much more restricted than in our model.   

\emph{Potentiation decay:} In addition to the activation threshold, 
a potentiation decay is applied to the weights of all
synapses, amounting to a slow memory leak within the system.
The decay is activity-independent and is implemented
as a rescaling of all synaptic weights $G_{i,j} \rightarrow \beta G_{i,j}$, where $0 < \beta < 1$,
as in previous phenomenological synaptic growth models
 \cite{kistler2002spike,grossberg1968some,toyoizumi2007optimality}. 
Long-term potentiation of synapses usually decays to baseline within three hours, 
which is called the early LTP (E-LTP)
 \cite{frey1997synaptic}. 
We assume that the reduction of
the synaptic weight during the trial time is insignificant given the typical three-hour time scale of potentiation
decay; therefore,
weight rescaling is applied between consecutive two-second training
trials, but not during the trial interval. 
Implementing the rescaling during trials is computationally intensive and produces no observable differences. 

In this simplified model of potentiation decay, all synaptic weights are subjected
to the weight rescaling between each trial, 
including weights of silent synapses.  
The decay of the silent synapses is important for our model. 
Consider what happens to synaptic weights that become deactivated due to either potentiation decay or synaptic depression. 
If synaptic depression were the only mechanism that modifies the weights of silent synapses, then 
a deactivated weight may remain close
to the activation threshold. This would lead to an
accumulation of weights near the threshold that require a small increase in order to become active.
On the other hand, if deactivated synapses have their weights immediately set to zero after deactivation, 
then a synapse that is consistently active, but does not evoke a spike from its target because of noisy fluctuations
over a few consecutive training trials, is immediately destroyed. Choosing to implement a decay
of the silent synaptic weights is proposed as a balanced solution to these two scenarios.
The decay of silent synapses can be related to gradual elimination of spines observed on dendrites \cite{holtmaat2009experience}.
Its biological mechanism is most likely different from the decay of E-LTP.
We apply the same decay rule for both silent and active synapses for the sake of simplicity. 
The details of how silent synapses decay do not matter. 
The functional role of the potentiation decay in our growth simulations is to regulate runaway synaptic growth.
We will demonstrate that, in combination, the activation threshold and the potentiation decay 
have a stabilizing effect on the growth of the network. 

\emph{Axon remodeling:} Synaptic weights are clipped if they are strengthened above a threshold
$G^{max}$ (see the Appendix). 
However, this does not limit the number of strong synapses that
approach the strongest allowed weight.
Another mechanism,
axon remodeling, regulates the number of strong synapses
a neuron can maintain with limited resources available. 
Limiting the number of strong synapses stabilizes an emerging synaptic topology of strong synapses.
If axon remodeling were not imposed on each neuron's axonal
tree, a well-connected neuron would continue to accrue targets.
Neurons \emph{in vivo} are observed sending out many axons during development,
then retracting most and maintaining only the strongest \cite{ruthazer2006stabilization}. 
A small number of strong synapses in a network have also been measured in experiments 
\cite{song2005highly,lefort2009excitatory}.
A slower potentiation decay is also applied to the strongest synapses, resulting in further stabilization.

Axon remodeling is implemented with the following rules, 
which are nearly identical to those in \cite{jun2007development}.
\begin{enumerate}
 \item A second threshold, $\Theta_S$, in addition to the active threshold, 
is introduced within the range of allowed synaptic strengths. 
Weights greater than this value characterize a strong active synapse, which we deem a \emph{supersynapse}.
Supersynapses elicit spikes reliably from a target despite the noisy fluctuations of the membrane potential.
The supersynapse threshold is greater than the active threshold: $\Theta_S > \Theta_A$.
\item A limit, $N_{ss}$, is imposed 
on the number of neurons that a presynaptic neuron contacts along supersynapses. 
This is the maximum number of axons a neuron can
maintain with its limited resources. When this number of supersynapses is attained, the neuron is said to be
``saturated.''
\item Once a neuron is saturated, the STDP rule is only applied to its supersynapses. After saturation
all synaptic weights continue to decay.
The potentiation decay reduces the weights of non-supersynapses, 
and as a result they will eventually approach zero with no opportunity to be potentiated
unless the neuron de-saturates.
\item Supersynapses are reinforced by repeated LTP;
without regular reinforcement, potentiation decay can cause de-saturation and all connections will undergo STDP again.
If de-saturation occurs frequently, no stable synaptic structure emerges.
High membrane potential variability
reduces the frequency of LTP at a supersynapse because higher noise reduces reliability of
a supersynapse to produce a spike from its target. In order to ensure LTP occurs frequently enough to overcome
the potentiation decay, in all
simulations, we apply a slower potentiation decay $\beta_{ss} = 1.1 \beta$ to supersynapses of a saturated neuron.
This corresponds to the slower decay of the late phase LTP (L-LTP) compared to E-LTP \cite{frey1997synaptic}. 
\end{enumerate}

Axon remodeling and the synaptic cap are non-essential to stability of the weight distribution of a network 
before training; they are only necessary when the network is presented with a stimulus. 
In the next section, the roles of
of axon remodeling will be further articulated
where the training regimen is described.

\subsection{Network training} \label{sec_training}

Network training is broken into a series of identical trials. 
Supersynapses can emerge within a network as it is presented repeatedly with a training stimulus. 
We model this stimulus with a short, strong excitatory current onto a small
subset of training neurons. 
The training excitation originates in an upstream brain area, possibly one processing sensory stimuli.
Training continues until the number of supersynapses
contained in the network stabilizes. 

A trial commences with the presentation of the training signal to the 
training neurons. The signal is modeled by a strong external drive
biasing the training neurons to spike within several milliseconds of the beginning of the trial.
After 8 milliseconds the driving current onto the training neurons 
returns to its baseline value.
The spontaneous activity and synaptic weight dynamics are simulated for one second after the training signal is
withdrawn.
The trial ends after this specified trial time 
and an inter-trial interval commences,
which we do not simulate. During this period, which is assumed much longer than one second, 
synaptic weights are reduced by the potentiation decay factor $\beta$ and the membrane potentials
are randomized. Training is repeated until the number of supersynapses reaches a stable value for 2,500 trials; 
this may take as few as 5,000 trials up to 100,000 trials depending on the size of the network and learning scale factor (see Appendix). 
We will show in the next section that
the training neurons form the
seed for development of a synaptic chain of neurons connected by supersynapses.

\section{Results}

\begin{figure}[!ht]
\begin{center}
\includegraphics[width=16cm]{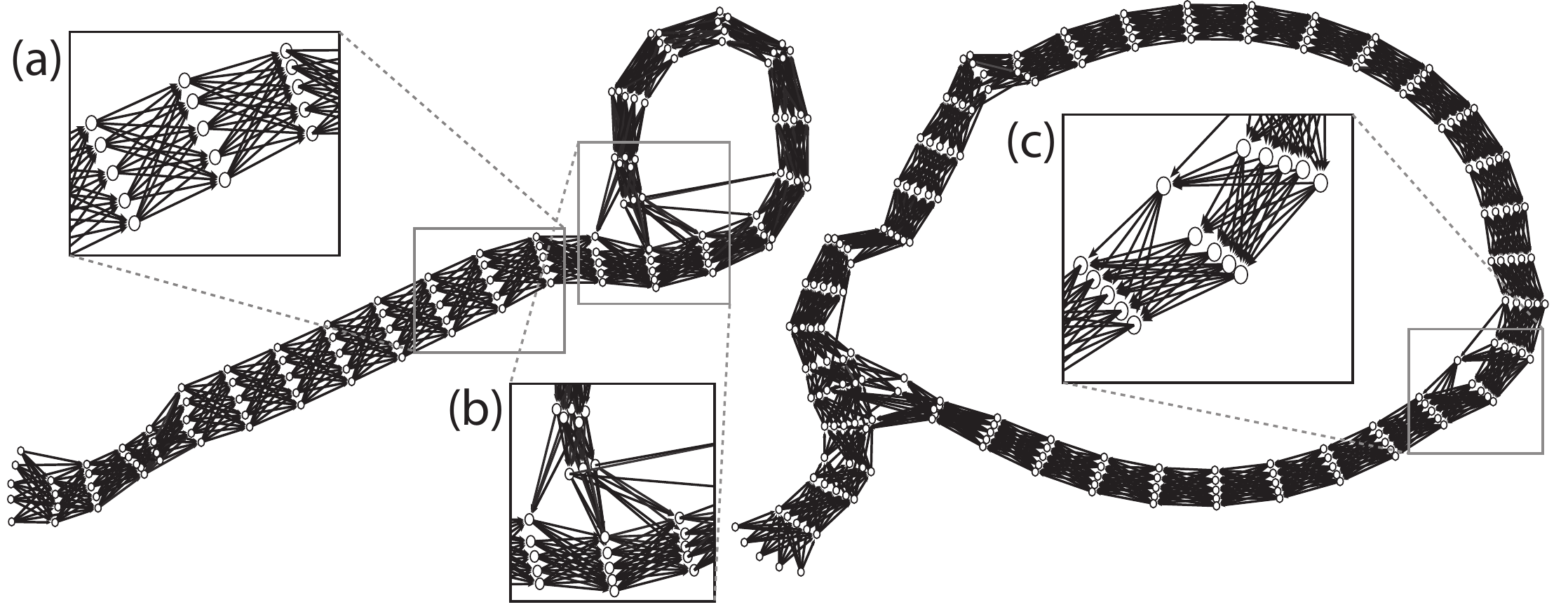}
\end{center}
\caption[Sample synfire chain.]{}
\label{chain_sam}
\end{figure}

\begin{figure}
\begin{center}
\includegraphics[width=8cm]{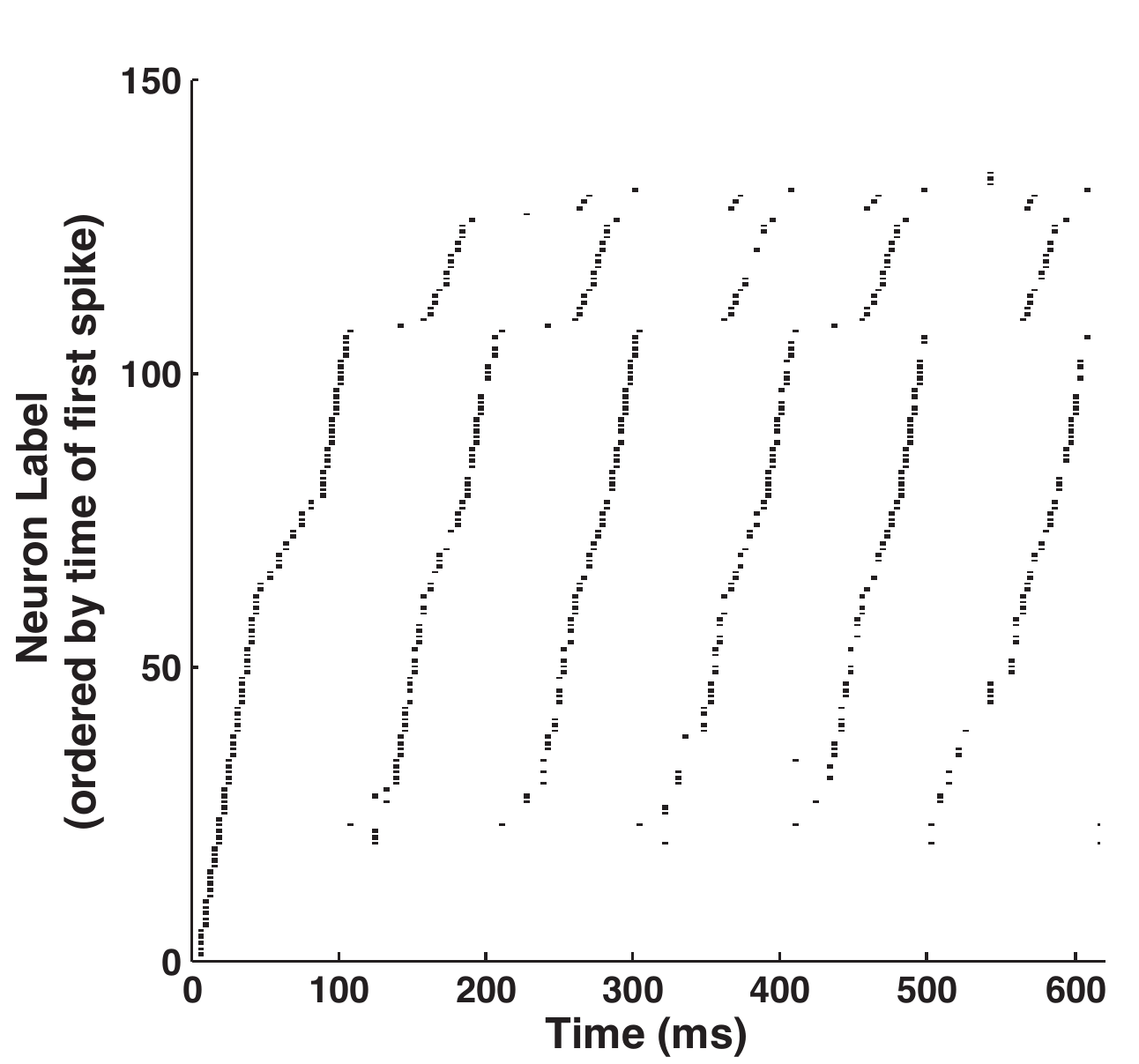}
\end{center}
\caption[Raster plot]{}
\label{raster}
\end{figure}

Synfire chain growth in response to training 
is governed by stochastic selection of post synaptic targets until 
a loop forms and the growth stops. 
Repeated stimulations of the training neurons 
change the synaptic weight distribution and produce strong, stable synaptic chain connectivity.
Chain growth emanates from the training neurons.
Neurons that spontaneously spike shortly after the training neurons may be targeted by the training neurons due to the STDP rule. 
Since the training neurons spike synchronously, they 
make convergent connections to the same set of neurons. 
Subsequent training strengthens these connections 
until the the synapses become supersynapses. 
Once supersynapses develop, reliable spikes can be evoked in these targets on nearly every trial. 
When this is the case, we say that the targets have been \emph{recruited}.
The cooperation via the convergent synapses is important for 
the targets to overcome membrane noise.
Axon remodeling restricts the number $N_{ss}$ of supersynapses that one neuron can maintain.
Consequently, the number of recruited neurons is close to $N_{ss}$ regardless of the number of the training neurons, although some fluctuations exist due to the noise in the recruiting process. 
The recruitment process continues as the second group accrues their own targets via the same cooperative process. 
New groups are recruited until previously-recruited neurons are recruited again, forming
a closed loop.
This stochastic, iterative process yields stable synfire
topologies that produce long, stereotypical sequences of spikes  \cite{jun2007development}. 
The chains consist of an introductory sequence that begins with the training neurons,
which feeds a loop of strong synaptic connectivity, examples of which are displayed in Fig. \ref{chain_sam}.
The network structure is clearly reflected in raster plots of the activity of the population 
during a typical trial after the chain is
fully formed, as shown in Fig. \ref{raster}.

The length of the chain formed by this process varies from trial to trial, and depends on the values of synaptic plasticity parameters and the size of the network 
(Fig. \ref{fig_synfire_sims}).
We find that the potentiation decay $\beta$ is the crucial model parameter that predicts the mean and variance of the
distribution.
We present a simple, analytically solvable, ``lottery'' chain-growth model 
to explain how the potentiation decay controls the characteristics of
the length distributions.
Our lottery model predicts
that the mean chain length approaches a finite value as network size is increased.
This simple model reproduces what is observed in the full, simulated model.
The mean chain length in the lottery model 
is controlled by a small parameter quantifying the likelihood that a neuron that is recruited by the chain
already targets the chain.
We observe these preferential connections from unrecruited neurons onto recruited neurons in the full simulations, and
we describe why these connections appear.
What these results show is that training  alters not only the network topology among neurons recruited to the chain, but also the connections from all other neurons to the neurons in a chain. 
This ``global'' response of the connectivity to an excitation targeting only a small subset of the population
is indicative of a synergistic relationship between the spike activity on the network and the underlying topology.

\subsection{Dynamical ground state}

Chain growth is initiated by 
stimulating the training neurons. Before training, 
the initial values of synaptic weights are drawn from a particular distribution, which we refer to as a \emph{dynamic ground state}.
Around 2-10\% of the synapses are active.
This connection probability of the active synapses is a generally accepted range for cortical networks \cite{braitenberg1998cortex}.
Spontaneous activity occurs in the network due to the noise and the active connections. 
The rules that govern synaptic dynamics
(STDP, active threshold, potentiation decay, etc.) 
yield a distribution of synaptic weights that is statistically stationary as the population is spontaneously spiking.
The dynamic ground state is stationary due to the homeostatic effect of the potentiation decay and the activation threshold.
If not for the interplay between these two plasticity rules, 
supersynapses would spontaneously emerge due to positive feedback
across the strongest synapses \cite{morrison2007spike}. 
Instead, a unimodal distribution of synaptic weights
emerges.
In Fig. \ref{necessary_decay} we compare a synaptic weight distribution when potentiation decay is acting on the
synaptic weights (Fig. \ref{necessary_decay}a) to when it is not (Fig. \ref{necessary_decay}b).
A particular synaptic weight in a dynamic ground state takes a random walk with steps generated by
the STDP rules and the potentiation decay.
Stability relies on potentiation decay that  
prevent synaptic weights from diffusing to large values. 
Synapses driven below the activation threshold by the potentiation decay and LTD can be reactivated by LTP. 
In a dynamic ground state, any neuron is connected to 
$\sim 2-10\%$ of the other neurons via active synapses at any given moment.

The distribution of synaptic weights in the dynamic ground state is obtained by
letting the weights evolve while simulating spontaneous activity without a training signal 
over a sufficent number of trials.
These ``initialization'' trials are identical to the training trials except that the 
training neurons are not subjected to the focused strong excitation.
Neurons are driven with noisy excitation over two-second trials
resulting in spontaneous activity while 
the synaptic weights evolve according to
the plasticity rules. After several hundred initialization trials, the stationary weight distribution emerges. 
We identify this network state when the number of active synapses in the network reaches a stable value. 
A dynamic ground state does not emerge for all sets of the synaptic plasticity parameters (details in the Appendix). 
For example, if the maximum possible potentiation of the synaptic weight is small compared to the activation threshold
and the potentiation decay is fast, 
the stationary state may contain only a few, short-lived active synapses
because newly activated synapses are driven below the threshold before they can be further strengthened. 
The opposite situation is also possible when the potentiation decay is too slow to admit a stable weight distribution
in which 2-10\% of the synapses are activated.
Finding the full parameter space for a stable dynamic ground state 
requires a parameter search.
However, a working combination can be found by 
setting the maximum potentiation slightly
larger than the activation threshold, and the potentiation decay rate 
fast enough to deactivate a newly activated synapse within 10's of trials.
These parameters produce a stationary distribution in which the number of active synapses is likely smaller than
2\%.
The number of  the active synapses can be increased 
by decreasing the maximum potentiation strength and 
the potentiation decay rate from this point. 

\subsection{Emergence of synfire chain}

When a training stimulus is presented repeatedly to a network in the dynamic ground state,
the stationary distribution of synaptic strengths is disturbed. 
This response of the network to the training stimulus
drives emergence of the synfire chain within the initially disordered ground state network.
During the training, 
the neurons recruited into the emerging chain 
have different synaptic strength distributions compared to 
those unrecruited (or ``pool'') neurons.
To illustrate why this is so, 
consider specifically the training neurons as they contact neurons in the pool.
Because the network is initialized in the dynamic ground state before training, all neurons, including the training neurons, have the same 
initial distribution of synaptic strengths onto their targets.  
However, when training begins,
the training neurons spike at the beginning of each trial with high probability, and the synaptic weights from the training neurons onto the pool neurons are more likely to increase because the training neurons
spike reliably every trial.
The weights approach a new equilibrium that has higher average weight 
than the dynamic ground state. 
This is shown in Fig. \ref{seq_shift}. The positive shift of the average strength of a synapse targeting the pool
is a result of spiking with near-certainty every trial. 

As the distribution of weights of synapses from training neurons onto the pool shifts positive, the 
potentiation decay
is not sufficient to deter rapid growth from positive feedback. Consequently supersynapses emerge
from the training neurons onto pool neurons. 
These synapses tend to be convergent since 
the convergence allows the training neurons to evoke reliably a spike from a shared target. 
Furthermore, the training neurons that do not share the target are
likely to develop connections onto a shared target since the shared target spikes frequently after the training neurons, which spike synchronously at the start of each trial.  
As the training progress, 
the training neurons accrue supersynapses onto shared targets, with their strengths capped at $G^{max}$.
When the number of supersynases from each training neuron hits the limit $N_{ss}$ imposed by the axon remodeling rule, all weak synapses are pruned and decay away due to the potentiation decay. 
Training neurons maintain only $N_{ss}$ supersynapses. 
Consequently, no more targets are recruited, and the second group is formed. 
The number of recruited neurons is close to $N_{ss}$ because of the 
convergence. 
Because of the strong, convergent connections from the training neurons, 
the second group spikes reliably in each trial. 
They accrue their own targets in the pool following the same process as the training neurons. 
The result is a positive shift of 
the distribution of synaptic weights away from the stationary distribution of the dynamic ground state.
Like the training neurons, the second group of 
neurons can eventually saturate by accruing shared targets within the pool
until axon remodeling prevents further growth.
The targets of the second group form a third group whose distribution of synaptic weights responds similarly.
Iterations of this recruitment process result in emergence of a synfire chain within the network.

As the chain network develops, 
spikes propagate along the chain when it is initiated by the training signal, and the ordering and timing of the spikes
is almost the same across trials.
A sequence may also be ignited
by spontaneous activity, which we call \emph{re-ignition}.
This can be observed directly in raster plots of spontaneous activity 
in networks with developing chains. 
An example is shown in Fig.  \ref{reignition_prob}a.
Spontaneous activity can initiate spike propagation from a random point in the chain. 
To quantify this observation, we simulated spontaneous activity 
of a network in which a subset of neurons are wired into a synfire chain and all other connections are randomly set (synaptic plasticity was suppressed).  
We measured the spontaneous firing rates of all neurons. 
As shown in Fig. \ref{reignition_prob}b,
the downstream neurons in the chain have higher firing rates than the upstream neurons.
This is because spikes reliably propagate down the chain wherever the re-ignition starts. 
The linear increase of the firing rates down the chain suggests that the probability of starting re-ignition is uniform across the chain. 

Re-ignition has direct impact
on the distribution of synaptic weights of the network.
After multiple re-ignitions, the number of neurons targeting the chain increases.
This is shown in Fig. \ref{reignition_prob}c. 
Pool neurons that are spontaneously active immediately before chain re-ignition have increased likelihood
of targeting the chain. Once these synapses from pool to chain are activated there is a 
decreased likelihood of LTD events on these synapses,
since the strong connectivity within the chain makes it more likely for activity to remain 
on the chain after chain neurons are spontaneously active. Hence, pool neurons tend to connect to a developing synfire chain.
This positive shift of the weights from pool neurons onto
neurons in the chain plays a role in the closure of the chain.
Once these preferential connections from the pool to the chain 
become numerous, it becomes likely that the pool neurons newly recruited into the chain are already connected to the chain, forming a loop that stops the chain growth. 
In Fig.\ref{reignition_prob}c it is clear that for faster potentiation decay, the total synaptic 
strength targeting the chain is smaller, implying that the stronger decay is more effective at reducing the strengths
of pool neurons targeting the chain. 
Chains recruit more groups and produce longer sequences if there are fewer pool neurons preferentially
targeting the chain, which can be facilitated by strong decay. 
The length distributions reflect this association (Fig. \ref{fig_synfire_sims}a): 
for slower potentiation decay, chains tend to be shorter with a smaller
variance, compared to chains subject to stronger decay. 

\subsection{Lottery growth model}

To test this association,
chain length distributions are generated by a simple lottery growth model. 
We model chain growth as a random process: 
neurons in the chain are drawn sequentially from a lottery of all neurons 
with equal probability. 
For simplicity, we assume that there is one neuron in each group in the chain, which is equivalent to setting 
$N_{ss}=1$ and using one neuron in the training set. 
The number of training neuron is also 1. 
At the $i^{th}$ iteration there are $i$ neurons in the chain 
out of the total network size $N$.
This simple model allows us to derive the chain length distribution analytically. 

We first consider the case that chain closes when a previously drawn neuron is re-drawn the second time, forming a loop in the chain and stopping its growth.
The probability that the ${(i+1)}^{th}$ neuron is drawn from the pool neurons and the chain does not close at length $i$  is 
\begin{equation}\label{cond_p_RR}
P(i+1|i)=\frac{(N-i)}{(N-1)} .
\end{equation}
The probability that it is re-drawn from the neurons in the chain and the chain closes at this iteration is 
$1-P(i+1|i)$.
Using these conditional probabilities, 
the probability $p_a$ of a mature chain with length $a$ is given by
\begin{equation} \label{gen_prob_a}
 p_a = \left[ 1-P(a+1|a) \right] \prod_{i=1}^{a-1} P(i+1|i),
\end{equation}
which, plugging in Eq. (\ref{cond_p_RR}), becomes
\begin{equation} \label{rand_recruit_stir}
 p_a \approx \frac{a-1}{N-1} \left(\frac{N-1}{N-a}\right)^{(N-a)} e^{-(a-1)}
\end{equation}
after applying Stirling's approximation.
To calculate the mean chain length as a function of network size $N$, we expand to lowest
order in $1/N$ and approximate the sum as an integral to find
\begin{equation} \label{RR_avgsize}
 \left< a \right> = \sum_{a=1}^\infty a p_a \approx
 1 - Ne^{-N/2} + \sqrt{2N} \int_{0}^{\sqrt{\frac{N}{2}}} dz e^{-z^2}.
\end{equation}
As $N \rightarrow \infty$, the mean chain length is on the order of $\sqrt{N}$ and is unbounded.
This is because the chance of re-drawing neurons in the chain is zero when $N \rightarrow \infty$. 

\begin{figure}
\begin{center}
\includegraphics[width=8cm]{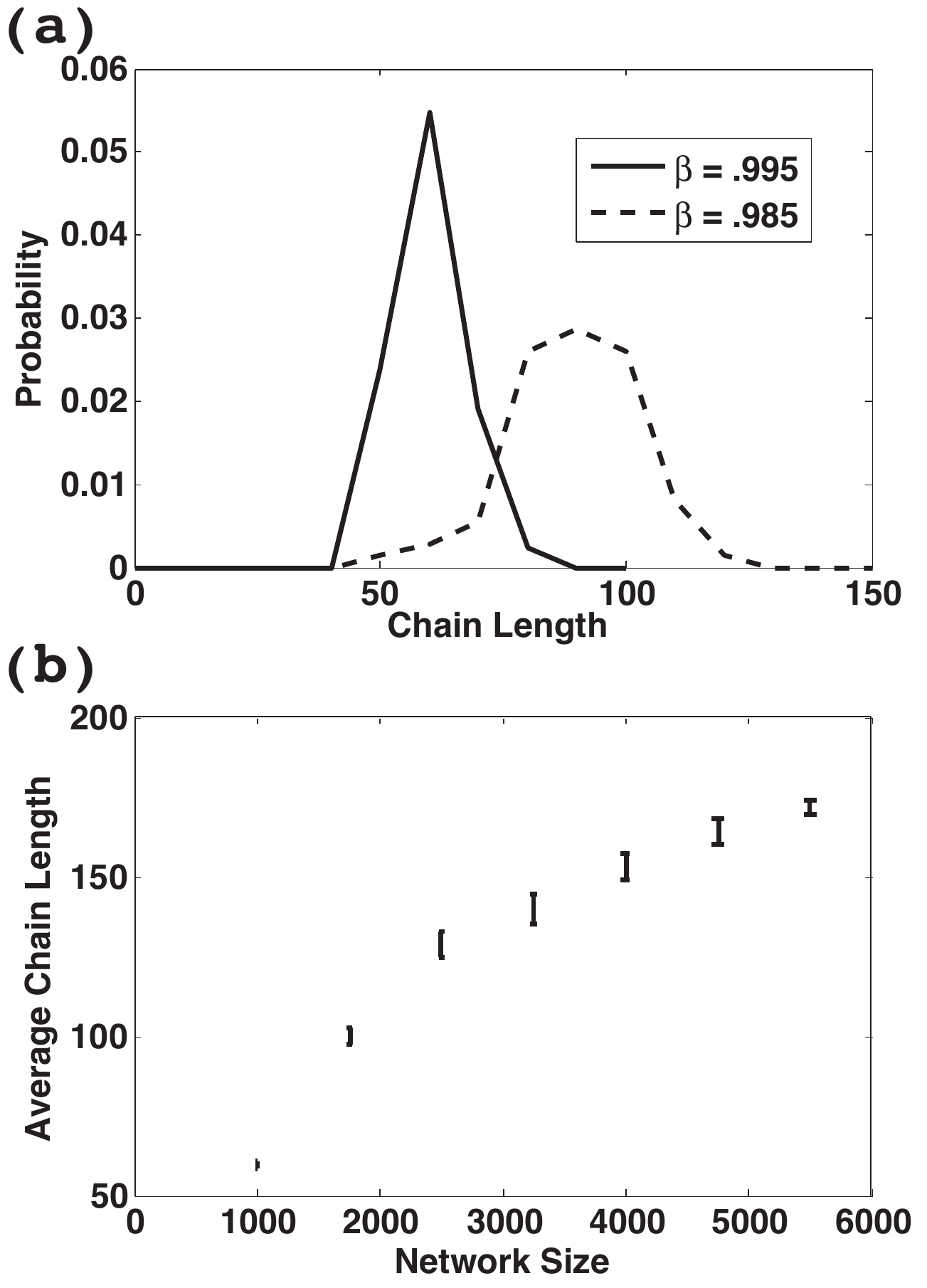}
\end{center}
\caption[Simulated chain length distributions with different rates of potentiation decay.]{}
\label{fig_synfire_sims}
\end{figure}

We now consider the case that chain also closes when a pool neuron preferentially connected to the chain is drawn, in addition to re-drawing a neuron in the chain. 
As we have shown in the previous section, slower potentiation decay enhances the probability of preferential 
targeting of the chain and reduces mean chain length (Fig. \ref{fig_synfire_sims}a).
To model this effect, we introduce a parameter $p_0$, which is the probability that a pool neuron is preferentially
targeting one neuron in the chain. 
The probability that the $(i+1)^{th}$ neuron is drawn from the pool and does not close the chain is 
\begin{equation}\label{cond_q}
 Q(i+1|i) = \frac{N-i}{N-1}(1-p_0)^{i}.
\end{equation}
There are two scenarios in which the chain ends with $a$ neurons. 
One, when the chain has length $a-1$, it can recruit
a neuron from the pool of $N-(a-1)$ neurons that has at least one connection onto a chain neuron. 
Two, when the chain has length
$a$, it can recruit one of the $a-1$ neurons above it in the chain. Therefore, the probability of chain closes at length $a$ has two terms:
\begin{equation}
\label{eqn-q-a-2}
 p_a = \left[ \frac{N-(a-1)}{N-1} - Q(a|a-1) \right] Q(a-1|a-2)...Q(2|1) + \frac{a-1}{N-1} Q(a|a-1)...Q(2|1).
\end{equation}
In the first term, the quantity in the brackets
is the probability of selecting a pool neuron that has at least one connection onto a chain neuron. 
Equation (\ref{eqn-q-a-2}) can be re-written into the form of Eq. (\ref{gen_prob_a}), with the conditional probability of chain not closing at length $i$ modified to 
\begin{equation} \label{cond_p_PA}
P(i+1|i)=\frac{(N-i)}{(N-1)} (1-p_0)^{i-1}. 
\end{equation}
Given the above conditional probability, the probability distribution of chain lengths is then:
\begin{equation}
\label{gen_prob_a2}
\begin{split}
 p_a = &\left[  (N-1)(1-p_0)^{-(a-1)} -(N-a) \right] \\
&\times p_a^{(RR)} \frac{(1-p_0)^{\frac{1}{2} a (a-1)}}{a-1} 
\end{split}
\end{equation}
where $p_a^{(RR)}$ is Eq. (\ref{rand_recruit_stir}), the probability of chain length assuming no preferential targeting. 

Equation (\ref{gen_prob_a2}) can be simplified in the large $N$ limit 
and moments of this distributions can be computed \cite{MillerThesis}. 
However, the expressions are too onerous to print here.
A notable feature is that the mean of this distribution approaches a finite limit as $N \rightarrow \infty$. 
In this limit, 
\begin{equation}
P(i+1 | i) = (1-p_0)^{i-1}, 
\end{equation}
and the probability of chain closing at length $a$ becomes
\begin{equation}
p_a = (1-p_0)^{(a-1)(a-2)/2} - (1-p_0)^{a(a-1)/2}, 
\end{equation}
according to Eq. (\ref{gen_prob_a}). 
The mean chain length is 
\begin{equation}
\left <a \right >_{N \rightarrow \infty}
= \sum_{a=1}^\infty a p_a
= 1 + \sum_{k=0}^\infty \left [ (1-p_0)^{1/2} \right ]^{(k+1)k}
= 1 + \frac{\vartheta_2(z=0,(1-p_0)^{1/2})}{2 (1-p_0)^{1/8}},
\end{equation}
where $\vartheta_2(z,q)$ is the Jacobi theta function \cite{thetafunction}. 
This is a finite number. 
Since every neuron in the pool has a non-zero probability of preferential targeting the chain, the mean chain length does not diverge even for $N \rightarrow \infty$. 

In Fig. \ref{fig_models}a we display several chain length distributions for different $p_0$. As $p_0$ is increased, the distribution shifts toward shorter chains and becomes sharper, indicating
that the chains close at an increasingly predictable length. 
This trend corresponds to the sharpening of the length distribution of synfire chains as the potentiation decay is slowed, shown in Fig.\ref{fig_synfire_sims}a. 

To confirm the model prediction that  the mean chain length approaches an asymptotic value even
as the network size grows very large, 
we performed a set of simulations with different network sizes.  
We set $N_{ss}=1$ and the number of training neurons to 1
to make the simulations directly comparable to the model.
Since neurons cannot cooperate, the variance of the GWN used in the simulations was reduced to $\sigma_V^2=1 \mbox{mV}^2$. 
Also the LTD time constant $\tau_D$ was set to $10$ms (see below). 
The potentiation decay was kept constant. 
The mean chain lengths in the simulations are well fit with the model using a single value of $p_0$, and show clear sign of saturation as the network size increases (Fig.\ref{fig_models}b). 
This trend is also observed in the fully complex, cooperative simulations 
which produce synfire chains.
Figure \ref{fig_synfire_sims}b is indicative of an upper bound on the length of the emergent synfire chain
as the network size is increased.

An minor effect omitted from the lottery growth model 
that also contributes to the shape of the length distribution is the
LTD window function (see Eq. (\ref{LTD}) in the Appendix).
The width of the window controlled by $\tau_{D}$ sets a soft minimum on sequence length. 
This effect was mentioned in
\cite{jun2007development}; here we show more detailed measurements in Fig. \ref{fig_LTD_shift}.
The effect can be attributed to reliable propagation of the training signal along a partially formed chain during each trial. 
A recruited neuron may target an upstream chain neuron directly 
or by targeting a pool neuron that is targeting the chain, contributing to the likelihood of chain closure. 
However, during each training trial upstream chain neurons 
spike before the newly-recruited neuron.
Therefore, the synapse 
onto the targeted neuron is weakened by LTD. 
If the temporal distance from the spike of the targeted neuron to that of the recruited neuron falls within the LTD window, 
the weight reduction quickly silences the synapse
and any possibility of reconnection is eliminated.
In the simulations that are used to validate the growth model (Fig. \ref{fig_models}), we used 
$\tau_D=10$ ms to minimize the LTD effect.  

Besides LTD, there are other simplifications
assumed by our growth model. 
A time-independent model parameter $p_0$ describes the probability that a pool neuron targets one of the entrained
sequence neurons and that the chain closes on itself by recruiting such a neuron.
This assumption ignores the non-equilibrium dynamics of the weight distribution as the chain recruits additional
neurons.
Re-ignition of the partial chain
precedes development of connections that preferentially target the chain. 
This is a random event that occurs at finite intervals, implying that preferential
connectivity has an associated time scale depending on the probability of a re-ignition event. 
The response time of preferential targeting can be seen directly in Fig. \ref{reignition_prob},
showing that the sum of weights targeting the chain, averaged over all members, saturates only after a number of
groups have formed. 
The above discussion of the effect of the LTD window function also indicates that $p_0$ is not uniform over the length
of the partial chain. 
In fact, it is effectively zero for neurons immediately upstream from the end of the chain.
Furthermore, because chain re-ignition is a random process driven by spontaneous activity, fluctuations
in the strength and number of synapses targeting the partial chain may contribute to the probability of closure.
A constant $p_0$ ignores such fluctuations. However, the model still gives reasonably accurate predictions. 

\section{Discussion}

In large recurrent networks with STDP, axon-remodeling and an activity-independent potentiation decay of synapses,
we observed emergence of long, stereotypical sequences
of spikes. The sequences are produced by stable synfire chain topologies that self-organize
via a stochastic growth process. We studied the distributions of synfire chain lengths and concluded that
the rate of potentiation decay in our synaptic plasticity model primarily controls the shape of the distributions.
The chains develop in response to a stimulus presented to the network
in a dynamic ground state,
in which the distribution of synaptic weights is invariant to synaptic modifications due to 
spontaneous activity on the network. This network state would not exist without
the potentiation decay. 

Synfire chain growth in our network model results from
a global response of the connectivity among the neurons to a stimulus that
targets only a small subset of the population, the training neurons.
Repeated stimulations of the training neurons leads to  
iterative growth of a synfire chain embedded in the network.
This result was expected based on previous work 
\cite{jun2007development}.
However, what was not expected, but what we observed, is global response of the connectivity 
as the chain develops. 
As the sequence begins to emerge, neurons in the pool are increasingly likely to target the neurons in the chain.
We suggest this process of targeting the strongly connected neurons in the chain is 
loosely analogous to preferential attachment in other complex networks
\cite{albert2002statistical}.
In contrast
to other systems with preferential attachment, a scale-free distribution does not emerge from training 
because of the topological constraint imposed on
the network by axon remodeling. 
The complex response of synapses throughout the network
illustrates co-evolution of spike activity (the emerging sequence) 
and synaptic topology (preferential targeting). We expect this observation 
generalizes to other recurrent network models with STDP in which spike sequences emerge. Since
pre-post synaptic strengthening is a common feature of STDP models,
other neurons will attach to sequence members when a sequence is initiated by spontaneous activity. 
We believe our insight may
explain the observation of neuron clustering \cite{cateau2008interplay,aoki2009co,levy2001distributed}
and small-world network degree distributions \cite{shin2006self,stam2010emergence} in other studies 
 where the number of strong connections
a neuron can make is unconstrained.
 
The coevolution of the network activity and network connectivity in response to an external stimulus
is reflected in the spectrum of length distributions of the synfire chains.
When the potentiation decay is too slow to sufficiently reduce the weights of connections
from pool neurons onto a partially formed chain, the variation of chain lengths is reduced.
When potentiation decay is fast, 
the number of preferential connections is reduced and the synfire chain has an opportunity to grow longer. 
We contrast our mechanism for synfire chain development 
with other studies in which chains emerge in a recurrent network, such as in Fiete, et. al. \cite{fiete2010spike}.
In Fiete's model, the synaptic plasticity rules are designed in such a way that each neuron (or group
of neurons receiving correlated external input) must connect to one other neuron (or neuron group)
that is not already targeted. The selection
of target is random, which leads to multiple closed loops where every neuron (or neuron group) is
incorporated into a loop. 
The distribution of chain lengths in this model follows a power law. 
Hence, short chains are more numerous than long chains. 
In contrast, the distribution of chain lengths in our model is close to a skewed Gaussian. 
There are typical chain lengths, and short chains are rare. 
In our model, not every neuron is part of the chain.
We introduced a growth model that incorporates preferential targeting
to confirm the general form of the length distributions of the chains. 
The growth model is verified with corresponding simulations of networks producing single-neuron chains.
The model illustrates tuning of the length distribution through the potentiation decay rate. 
Furthermore, it predicts that the mean chain length
approaches a constant in the limit of large network size. Simulations of the more complex process
of synfire chain growth confirm the same saturation effect (Fig. \ref{fig_synfire_sims}).
This is in contrast to the case of $p_0=0$, for which the mean
length diverges as $\sqrt{N}$.
Any small preferential targeting probability $p_0$ limits the mean length as $N \rightarrow \infty$.
This result indicates that chain size is bounded softly, even in the limit of very large networks.
It would be interesting to confirm this plateau effect in recurrent networks larger 
than those we were able to simulate.
In at least one case we know of \cite{kunkel2010limits}, a much larger network has been simulated.
However, chains did not emerge upon externally stimulating the network in this study.
Contrasting the result of this study with our own, 
we have validated the iterative recruitment of synfire chain groups 
using a power-law STDP rule \cite{morrison2007spike} instead of the additive LTP/multiplicative LTD model 
 (see Eqs. (\ref{LTP}) and (\ref{LTD})) introduced
by \cite{vanrossum2000stable}.
Additionally, we observe the growth process is unaffected when setting the number of allowed strong connections to larger values (50 instead of 
5 used in the simulations presented in the Results), 
demonstrated also in \cite{jun2007development}. 
Key differences that may account for the emergence of chains in our model are, dually, the vast restructuring
of the network connections allowed by imposing an activation threshold on each synaptic weight,
and also restricting the influence of a single neuron by imposing the axon-remodeling rule. 

Before training, networks are initialized to a dynamic ground state. The distribution of synaptic weights
in a ground state network is stationary while neurons are spontaneously active.
Synapses are activated and silenced by random activity, 
and the average flux of weights across the active threshold is zero.
Our synaptic dynamics model is distinguished from others in two ways. One, we impose an activation threshold
on a synaptic weight between every pair of neurons in the network. 
The picture that emerges is one in which
neurons are actively connecting and disconnecting to other neurons in the 
population freely and on a relatively 
short time scale, minutes to hours. A number of imaging studies 
support this fast restructuring of network connectivity patterns \cite{holtmaat2009experience}. 
The time scale of emergence and subsequent withdrawl of dendritic spines 
can be as short as ten minutes and has been linked to synaptic activity \cite{maletic1999rapid}.
Network rewiring is not permitted in all but a few network growth models that have been proposed 
\cite{bamford2010synaptic,iglesias2005dynamics}. Instead, it is much more common to select \emph{a priori}
the postsynaptic targets of each neuron. We argue that this modeling choice neglects an important feature of
biological networks and places limits the emergent topology of the network.
Two, we subject all synapses to an activity-independent decay. 
We propose that this is related to the widely observed decay of the 
early phase long-term potentiation (E-LTP). The time scale of the 
potentiation decay is several hours \cite{frey1997synaptic}, much longer than the length of an individual training trial.
The role of the potentiation decay is to avoid the accumulation of random potentiation of synaptic
weights known to destabilize the network dynamics \cite{morrison2007spike}.
The combination of these two rules yields a robust spectrum of stationary network states.
As a final note, we compare our network rewiring rules with a similar approach taken by Iglesias et al
\cite{iglesias2005dynamics}.
In this study a recurrent network is initialized with all-to-all connectivity
and network connections are eliminated via STDP. This modeling choice is also notable in that
final network connectivity is not limited by only modifying weights between specific pairs of neurons.
However, it is unclear from the results of this study whether sequences emerge after pruning.

The stability of the ground state network indicates the rules of our model
encode a homeostatic mechanism \cite{turrigiano2004homeostatic}. 
Several other models of homeostatic mechanisms have been proposed recently, including
a sliding modification threshold based on post-synaptic firing history \cite{benuskova2007stdp},
a dependence on fluctuations of post-synaptic membrane potential \cite{clopath2010connectivity}
and heterosynaptic plasticity that limits the total weight targeting a single neuron \cite{fiete2010spike}.
An overlooked mechanism that we propose is activity-independent, multiplicative rescaling of weights, 
as we have implemented here, potentiation decay or decay of E-LTP.
This form of LTP returns synaptic efficacies to the baseline within 3 hours 
\cite{frey1997synaptic} and is independent of protein synthesis. 
Only through repeated potentiation, E-LTP can turn into the late phase LTP (L-LTP), 
which is maintained by protein synthesis and can last over days and weeks \cite{frey1997synaptic}. 
In our model, consistently potentiated synapses turn into supersynapses whose decay is much weaker 
than other active synapses. 
The supersynapses can be considered in the L-LTP state. 
Emergence of the synfire chain relies on stabilization of the small percentage of supersynapses, 
while there are many
weaker, more transient synapses. This long-tail synaptic weight 
distribution is consistant with physiological observation \cite{song2005highly,lefort2009excitatory}
and appear in other theoretical studies \cite{teramae2012optimal,gilson2011stability}.
The functional role of E-LTP decay is largely ignored in the LTP literature. 
Our model suggests that the E-LTP decay may be crucial in stabilizing synaptic weight distribution against random accumulations of LTP through spontaneous activity. 
Moreover, the E-LTP decay can be important in the formation of functional networks through STDP. 
The time scale of the potentiation decay in our model is congruent with time scales of E-LTP decay,
which can be seen with an order of magnitude estimate.
If we assume learning occurs on a scale of tens of days and $10^4$ training trials are necessary for synaptic chains to crystalize,
this places the time scale corresponding to 
our decay parameter $\beta$, which we vary from 0.9 to 0.99, in a range of $10^3$ and $10^4$ seconds. 
It will be interesting to test these ideas by manipulating the E-LTP decay constants in developing or learning brains {\it in vivo}. 

A natural extension of this work is to construct a growth model in such a way that more complex asymptotic
synaptic topologies emerge. 
Many learned motor behaviors can be complex. For example, 
stochastic ordering of distinct elements of a behavior is one kind of 
behavioral complexity. The song of the Bengalese finch can be described by this type of 
stochastic process \cite{jin2009generating,jin2011compact}. 
A single synfire chain cannot capture this complexity
since they produce only a single spike sequence; multiple chain or branching chains would be necessary. 
One possible scenario for growing multiple chains in the same network would be to have distinct sets of training neurons.
However, it turns out that preferential targeting of sequence members
within the network prohibits the development of
distinct synaptic structures.
We implemented two distinct training groups of 5 neurons in a network of 2500 neurons.
A training neuron set that is excited at the beginning of each learning trial.
Which of the two sets is selected at random with equal probability.
In Fig. \ref{two_training} we display the resulting growth.
The chains develop several groups individually, but ultimately they merge to a single chain.
Initially, the training neurons seed two disjoint sequences that recruit targets iteratively. Emergence of synfire
groups embeds two disjoint sequences in the network. 
These sequences are occasionally activated by spontaneous activity. Therefore,
neurons in the pool will target the partial chains preferentially. In particular, neurons at the end of one
of the chains may target a neurons in the other chain with elevated probability. Once one chain reliably
activates neurons in the other chain, they will merge. Merging occurs reliably each time we simulated two
training groups. Other growth mechanisms must be present, or the chains must be encoded within distinct populations
of neurons. This conclusion is consistent with other studies, such as \cite{verduzco2012model}.

\section{Conclusions}

In a recurrent network of neurons driven by high-frequency noisy input and synapses governed by a set
of plasticity rules which include STDP, a potentiation decay and axon remodeling,
we showed that neurons cooperate via convergent synapses to self-organize into a synfire chain
characterized by a precisely-timed sequence.
The network is initialized in a state characterized by
a statistically stationary distribution of synaptic weights, invariant to network spontaneous activity.
The combination of a potentiation decay plus an activation threshold imposed on the synaptic weights
provides a homeostatic mechanism within the network.
A small subset 
of neurons stimulated by a strong excitation forms the seed for recursive synaptic growth of synfire groups.
During repeated presentations of this stimulus and emergence of the chain, 
we observe a complex response 
of the network connectivity that is reflected in the distribution of asymptotic chain lengths.
We have demonstrated a clear example of interplay between neural activity and emergent synaptic
topology in a developing network.

\section{Appendix}

\subsection{Neural dynamics}

The simulated networks consist of $N$ excitatory, conductance-based, 
pulse-coupled leaky integrate-and-fire (LIF) neurons.
The state of the $i^{th}$ neuron is described by a single variable $V_i$, its membrane potential, which obeys
\begin{equation} \label{LIF_eq}
\tau \frac{dV_{i}}{dt} = L - V_{i}(t) + S_i(V_i,t), 
\end{equation}
where $S_i(V_i,t)$ is synaptic input to the membrane. The LIF neuron requires several parameters:
leak reversal potential $L=-70 \mbox{mV}$, membrane time constant $\tau=20 \mbox{ms}$, spike threshold 
$\theta=-54 \mbox{mV}$ and reset potential $V_{R}=-70 \mbox{mV}$. If $V_i = \theta$, the neuron emits
a spike and is instantaneously reset to $V_{R}$.

The synaptic input to the $i^{th}$ 
neuron $S_i(V_i,t)$ consists of three sources: a noisy external drive, an excitatory
conductance, and an inhibitory conductance:
\begin{equation}
\begin{split}
 S(V_i,t) = &\left( I_{ext} + \xi_i(t) \right) + g_i^{(E)}(t) \left[E^{(E)} -V_i \right] \\ &+ 
g_i^{(I)}(t) \left[ E^{(I)} -V_i \right].
\end{split}
\end{equation}
We choose the reversal potentials $E^{(E)}=0 \mbox{mV}$ and $E^{(I)}=-75 \mbox{mV}$.
The drive includes a Gaussian white noise (GWN) term $\xi_i(t)$, obeying $\left< \xi_i(t) \right> = 0$
and $\left< \xi_i(t)\xi_i(t')\right> = \sigma_V^2 \delta(t-t')$ with all higher order correlations equal to zero.
The noise is uncorrelated across individual neurons.
Driving current is $I_{ex}=25 \mbox{mV}$ and $\sigma_V^2 = 10 \mbox{mV}^2$.
Training neurons (see Methods) are subjected to larger driving current (100 mV) for the first 8ms of each training trial.
The external drive originates in upstream regions, which we do not simulate.
Gaussian white noise is commonly employed to model this input \cite{lansky2006parameters}.

The conductances $g_i^{(E)}(t)$ and $g_i^{(I)}(t)$ take the form of sums of $\delta$-functions
centered on the spike times of neurons in the network.
Specifically,
\begin{equation}\label{ex_cond}
 g^{(E)}_{i}(t) = \tau \sum_{j=1}^{N} \sum_{p=1}^{\infty} G^{(E)}_{j,i} \delta(t-T^{(j)}_{p}),
\end{equation}
where $G^{(E)}_{j,i}$ is the excitatory synaptic weight from neuron $j$ onto $i$
and $T^{(j)}_p$ is the time of the $p^{th}$ spike of neuron $j$.
Weight $G^{(E)}_{j,i}$ is zero if $j$ does not have a synapse onto $i$. 
$G^{(E)}$ is in a range $[0,.275]$, expressed as a multiple of the neuron's leak conductance. 
Also, we forbid self-synapses: $G_{i,i}=0$.

\subsection{Global inhibition}

Global inhibition, i.e. inhibition that targets all neurons, 
is concurrent with each spiking event in the network. The inhibitory conductance takes the form
\begin{equation}\label{inh_cond}
 g^{(I)}_{i}(t) = \tau \sum_{j=1}^{N} \sum_{p=1}^{\infty} G^{(I)}_n \delta(t-T^{(j)}_{p}).
\end{equation}
Here $G^{(I)}_n$ is the inhibitory conductance on each excitatory neuron induced by $n$ nearly synchronous (within the time resolution $T_{res}$, Methods, section A) spikes of the excitatory neurons at any single iteration of the population dynamics.
$G^{(I)}_n$ is computed under the following assumptions: 
1) the feedback inhibition is mediated by $N_I$ inhibitory neurons; 
2) each excitatory neuron randomly connects to $k$ inhibitory neurons; 
3) when excited, an inhibitory neuron emits a single spike with certainty regardless the number of nearly synchronous spikes it receives. 
The maximum inhibition that can be provided to an excitatory neuron is $N_I G^{(I)}_1$, where
$G^{(I)}_1$ is the weight of a single inhibitory synapse.
We can derive that 
\begin{equation}\label{inh_series}
 G^{(I)}_{n} = G^{(I)}_{1} s(n) = 
G^{(I)}_{1}   \sum_{i=0}^{n-1} k \left(1-\frac{k}{N_I} \right)^i.
\end{equation}
Here $s(n)$ is the average number of inhibitory neurons contacted by $n$ excitatory neurons. 
Each term in the sum over $i$ in Eq. (\ref{inh_series}) 
is the average number of addition interneurons contacted by the $(i+1)$-th excitatory neuron. 
Note that $(1 - k/N_I)^i$ is the probability that an inhibitory neuron is {\it not} yet contacted by previous $i$ excitatory neurons. 
For all simulations, we pick $k/N_I=.05$
and $G^{(I)}_1 k = 0.8$.

\subsection{Simulation algorithm}

The computational advantage of using pulse-coupled neurons is that the response of each neuron to a spike is instantaneous and
can be calculated by plugging Eq. (\ref{ex_cond}) and Eq. (\ref{inh_cond}) into Eq. (\ref{LIF_eq}) and integrating
over the $\delta$-functions in an infinitesimal neighborhood $\epsilon$ around spike time $T_j^{(p)}$ \cite{jin2002}.
The result is
\begin{equation} \label{MP_resp}
\begin{split}
 V_i(T_j^{(p)}+\epsilon) = &e^{-(G^{(E)}_{j,i}+G^{(I)})} \left[ V_i(T_j^{(p)}-\epsilon) 
  -\frac{G^{(I)} E^{(I)}}{G^{(E)}_{j,i}+G^{(I)}} \right]
    \\ &+ \frac{G^{(I)} E^{(I)}}{G^{(E)}_{j,i}+G^{(I)}}
\end{split}
\end{equation}
as $\epsilon \rightarrow 0$. Since the interactions are instantaneous, 
between spikes the probability distribution $\rho^{MP}$ describing the time-evolution
of the membrane potential is the solution of a Fokker-Planck equation \cite{capocelli1971diffusion}
\begin{equation}\label{SLIF_fp}
\begin{split}
 \frac{\partial \rho^{MP}}{\partial t} (v,t;W) = 
&-\frac{\partial}{\partial v} \left[ \frac{L+I_{ex}-v}{\tau} \rho^{MP}(v,t;W)
\right] 
\\ &+ \frac{\sigma_V^2}{\tau} \frac{\partial^2 \rho^{MP}}{\partial v^2}(v,t;W),
\end{split}
\end{equation}
where the variance of the distribution in the large $t$ limit $\sigma_V^2=10 \mbox{mV}^2$. 
$W$ is the initial value of the membrane potential.
An fast event-driven algorithm \cite{reutimann2003event} 
has been developed to generate spike activity on networks of pulse-coupled neurons with GWN
using ``lookup tables'' containing the solutions to Eq. (\ref{SLIF_fp}), as well as distributions of 
spike threshold first passage times
\cite{tuckwell1989} describing spike-timing distributions.

The algorithm has four steps. 
\begin{enumerate}
\item Calculate a predicted spike time for each neuron, given an initial
value of the membrane potential $W$, by sampling the first passage time distributions.
\item Identify the neuron with the minimal predicted spike time. Also, identify any neurons that is predicted
to spike within $T_{res}=3 \mbox{ms}$ of the neuron with the minimal predicted spike time.
These neurons will be next to spike.
\item Assign a membrane potential to each neuron that does not spike by sampling the membrane potential
distributions at the spike time calculated in Step 2. Reset the neurons that spike to $V_R$. 
\item Apply Eq. (\ref{MP_resp}) to determine the membrane potentials immediately after the spike is applied.
\item Calculate new values of synaptic weights according to STDP and other synaptic dynamics rules.
\end{enumerate}
These steps are iterated as long as desired.

\subsection{STDP window}

The weight of a synapse is updated according to an STDP rule during the last step of the simulation algorithm.
The weight of a synapse from excitatory presynaptic neuron $i$ to postsynaptic neuron $j$ is $G_{i,j}$.
We implement the additive/multiplicative rule \cite{vanrossum2000stable}. We introduce also a tunable 
scale factor $R$ \cite{guetig2003learning}. The weight is modified using the following STDP window:
\begin{equation} \label{LTP}
G_{i,j} \to G_{i,j} + R A_{P} G_{P} F_P(|\Delta t|)
\end{equation} 
\begin{equation} \label{LTD}
G_{j,i} \to G_{j,i} - R A_{D} G_{j,i} F_D(|\Delta t|)
\end{equation} 
with $\Delta t = T^{(j)}-T^{(i)}$, the spike time difference, $A_P=.01$, $A_D=.0105$, and unless otherwise specified
$G_P=.1$.
We impose $R A_D < 1$ to ensure non-negative weights. Weights are clipped above $G^{max}=.275$.
The long-term potentiation(P)/depression(D) function is defined as
\begin{equation} \label{stdpfn}
F_{P(D)}(t) = \left\{ \begin{array}{ll}
(1 \pm f_0)t/t_{P(D)} \mp f_0 & \textrm{if $t<t_{P(D)}$} \\
e^{-(t-t_{P(D)})/\tau_{P(D)}} & \textrm{if $t \geq t_{P(D)}$}
\end{array} \right. 
\end{equation}
This rule is identical to the rule proposed in the Jun-Jin model \cite{jun2007development} except
for the introduction of an additional parameter $f_0$.
For nonzero $f_0$, 
there is a net reduction in weight for $\Delta t = 0$ \cite{babadi2010intrinsic}, the magnitude of which is
controlled by the parameter $f_0$; we pick $f_0=.25$.
This rule has been shown to be equivalent to the effect of jitter on the arrival times of pre- and post- spikes at the synapse
and compliments the time resolution we impose on the spike arrival times.  
Functionally, it discourages connections between neurons that spike reliably within $T_{res}$
and effectively reduces any the weights of extraneous active synapses 
that may exist among neurons recruited to the same synfire group. 

\section{Acknowledgement}

This research was supported by the Sloan Research Fellowship, the Huck Institute of Life Sciences at the Pennsylvania State University, and NSF Grant No. IOS-0827731. 
We thank Jason Wittenbach for his comments on the manuscript.


%

\newpage
\section*{Figures}

\noindent
FIG.1. 
{Samples of asymptotic configurations of synapses in a 4000 neuron network generate long, stereotypical
sequences of spikes. The chains above have 25 (left) and 40 (right) synfire groups. The time between firing
of adjacent groups is $\sim 3-4$ ms. 
The number $N_{ss}$ of super synapses one neuron can maintain is set to 5. The number of training neurons is 5. 
{\bf (a)}
Blowup depicts regular synfire connectivity between groups. 
The regular structure observed is a result of neurons cooperating to excite targets.
A neuron most effectively excites a target when the target is shared with a neuron in its group,
so convergent synapses are favored for development into supersynapses.
A group accumulates shared targets until the maximum allowed number of supersynapses is reached.
There are five neurons per synfire group in this
network because the maximum number of supersynapses allowed by axon remodeling is five per neuron. 
{\bf (b)} The reconnection point is splayed across several groups. The connections that
form first are to the group nearest the top of the chain. 
The downstream connections follow due to elevated probability of re-ignition at the initial connection point.
The splayed connectivity allows spontaneous activity to restart when the excitation reaches this point because
neurons across several synfire groups spike and sequential activity is most stable when a full synfire group fires.
However, the synapses at the reconnection point potentiated often enough to remain stable.
{\bf (c)}
Defects sometimes appear as the chain emerges. Small defects like the one depicted 
can remain stable. Severe defects are not stable
and never appear due to lack of a clear sequence that is consistently reinforced by STDP.}\\

\noindent
FIG.2. 
{A raster plot of the population activity after a synfire chain has self-organized. The neurons are labeled
according to their time of first spike. Some neurons in the chain do not spike during the first iteration
around the loop
due to fluctuations of the membrane
potential, but they do spike during subsequent iterations. These neurons have the highest valued labels
across the top of the plot.
}\\

\noindent
FIG.3. 
{{\bf (a)} Two simulated synfire chain length distributions for two different potentiation decay parameters.
As the potentiation decay slowed, the resulting distribution of lengths narrows. 
Sample size: 100 networks. $N = 1000$ neurons.
{\bf (b)} The mean synfire chain length as a function of network size. 
The mean length saturates as the network size increases.
Sample size of each data point: 100 networks. $\beta = .985$. 
The error bars denote the standard error of the mean.
}\\

\noindent
FIG.4.
{The potentiation decay suppresses runaway
synaptic growth resulting from positive feedback along active synapses. 
Two networks are compared by setting 
the activation threshold to $.015$ (``active'') and $\infty$ (``silent''), respectively.  
The ``silent'' network effectively has all synaptic conductance set to zero.
In {\bf (a)}, a potentiation decay is applied to all synapses after each 1 s interval of simulated time. In {\bf (b)},
there is no potentiation decay.
In both scenarios there is no positive feedback in the silent networks, 
so these distributions (light gray) are stable in both {\bf (a)} and {\bf (b)}.
However, in the networks with active synapses, only the distribution in the network with potentiation decay {\bf (a)}
remains stable indefinitely. In the network without potentiation decay {\bf (b)}, synapses grow large resulting in
runaway activity.}\\

\noindent
FIG.5.
{Histogram of synaptic weights in network with silent synapses.
A group of training neurons spikes at the beginning of every trial.
The distribution of weights onto the non-training neuron (pool neurons) stabilizes with a higher mean weight. 
The synapses in these simulated networks are silenced (i.e., $\Theta_A > G^{max}$) in order to emphasize
that the net strengthening is independent of interactions between individual neurons.
When interactions are allowed, the strongest synapses may overcome the potentiation decay leading to development
of strong synapses within the network.}\\

\begin{figure}
\begin{center}
\includegraphics[width=8cm]{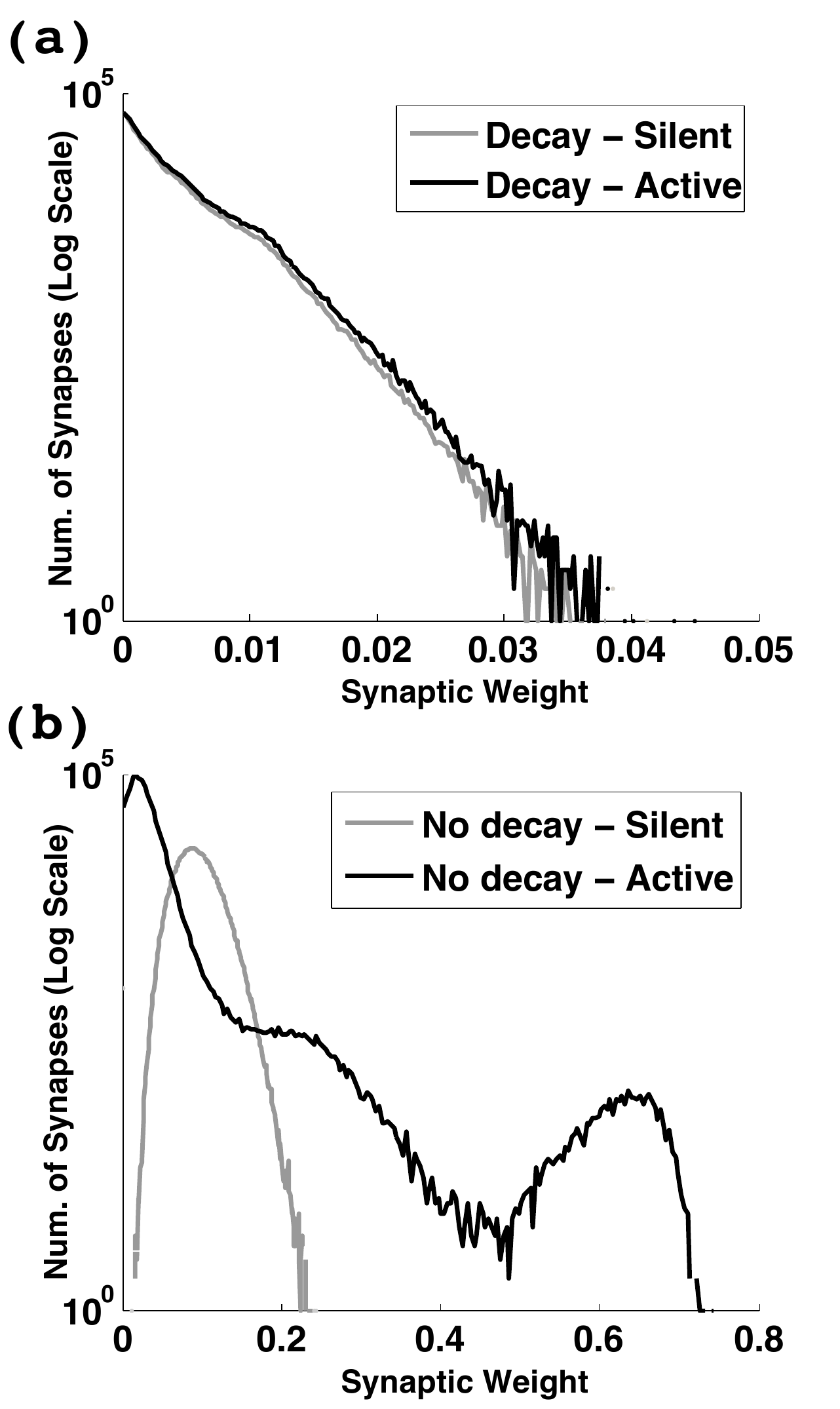}
\end{center}
\caption[Potentiation decay and unstable synaptic weight distributions]{}\label{necessary_decay}
\end{figure}

\begin{figure}
\begin{center}
\includegraphics[width=8cm]{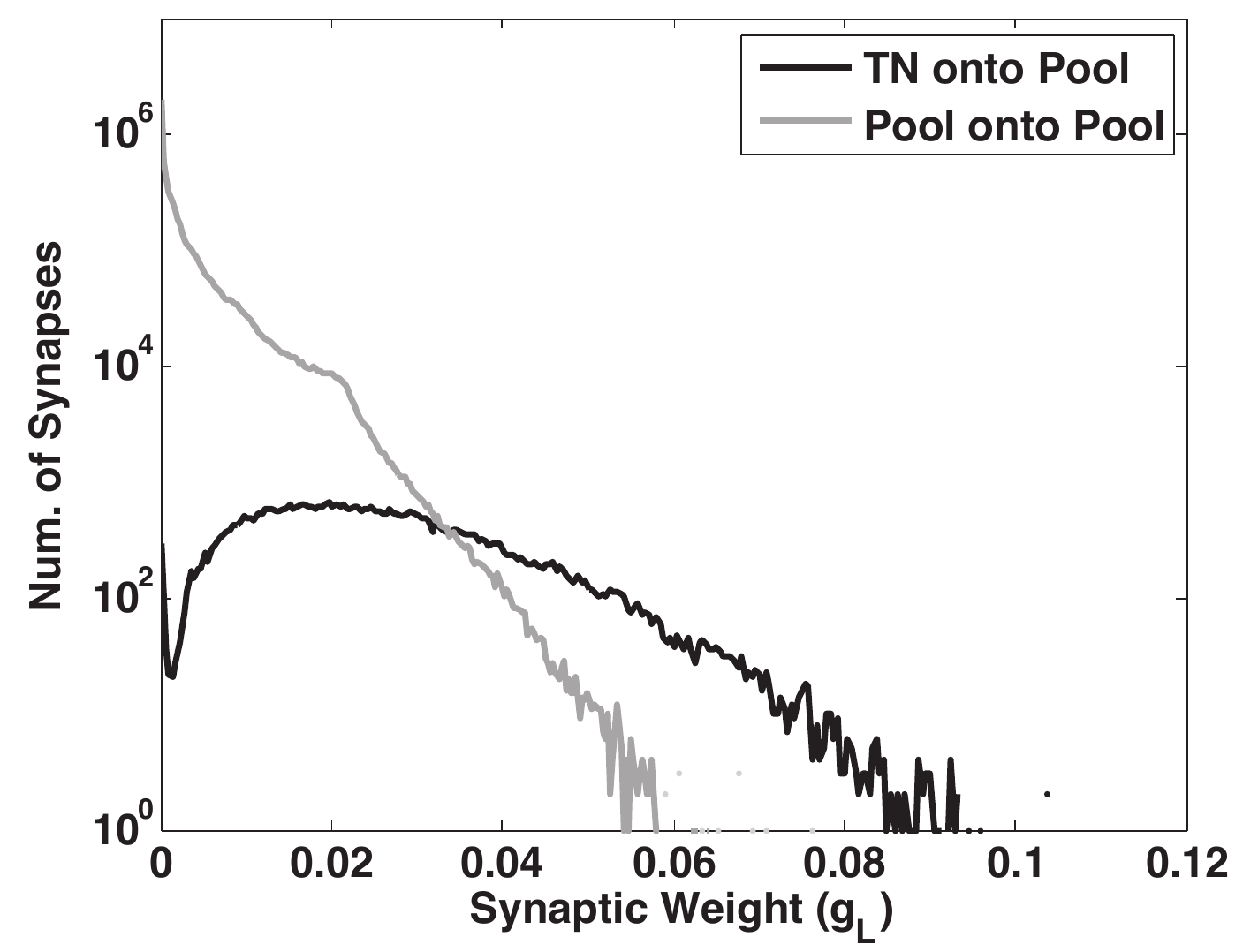}
\end{center}
\caption[Embedded sequences in the network produce different equilibrium distributions.]{}
\label{seq_shift}
\end{figure}

\begin{figure}
\begin{center}
\includegraphics[width=8cm]{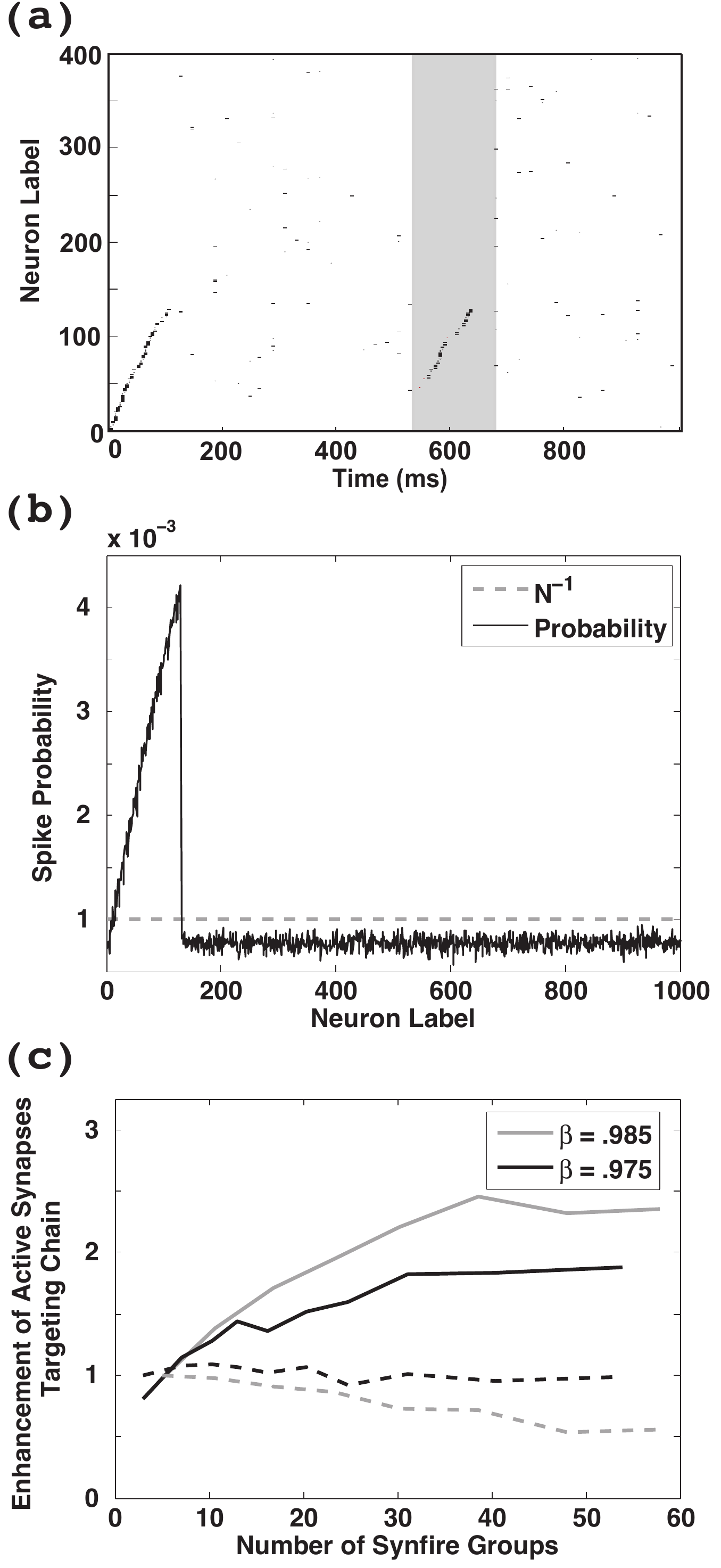}
\end{center}
\caption[Spike probability enhancement for sequence members.]{}
\label{reignition_prob}
\end{figure}

\begin{figure}
\begin{center}
\includegraphics[width=8cm]{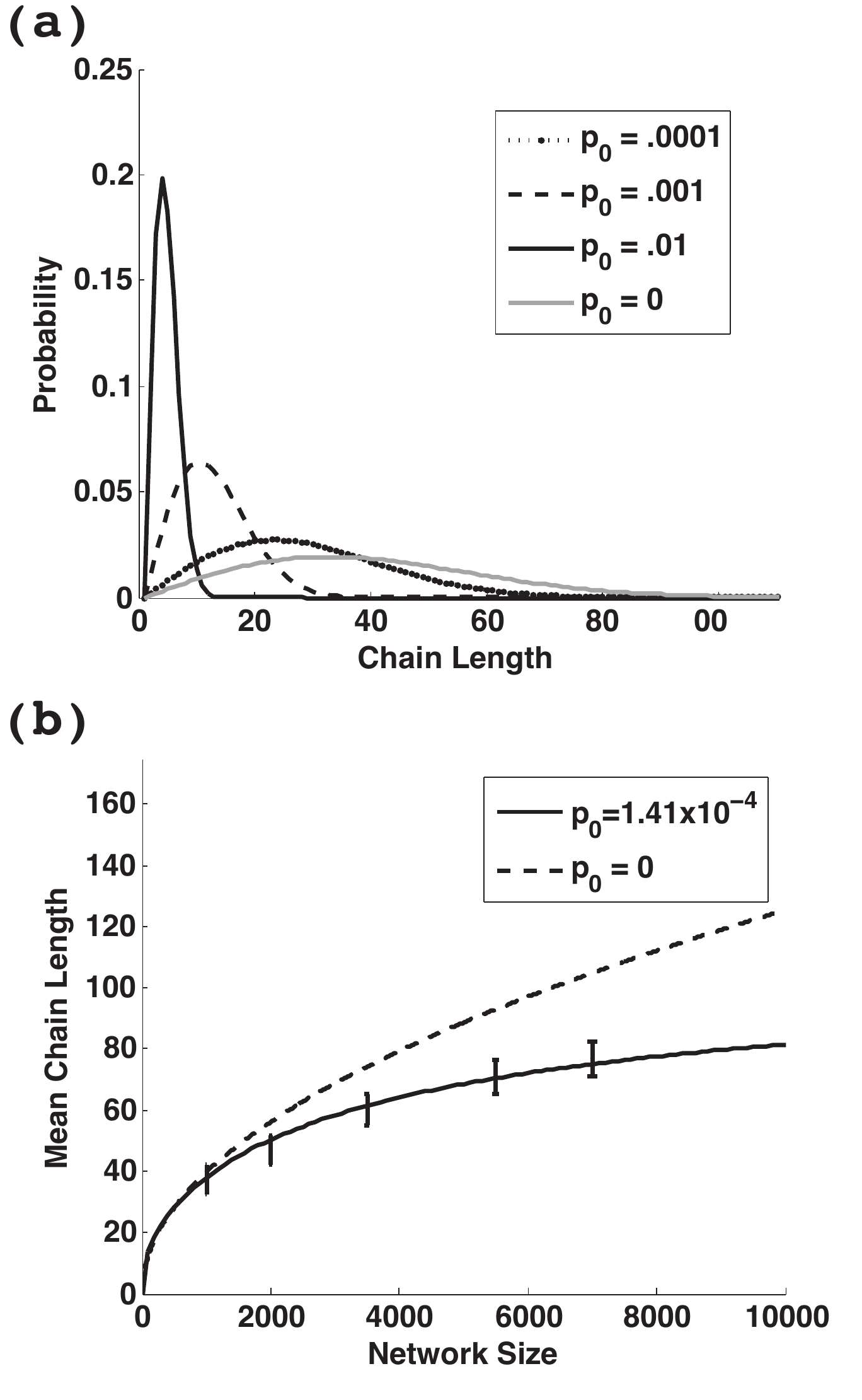}
\end{center}
\caption[Simulated distribution and comparisons with growth models.]{}
\label{fig_models}
\end{figure}

\begin{figure}
\begin{center}
\includegraphics[width=8cm]{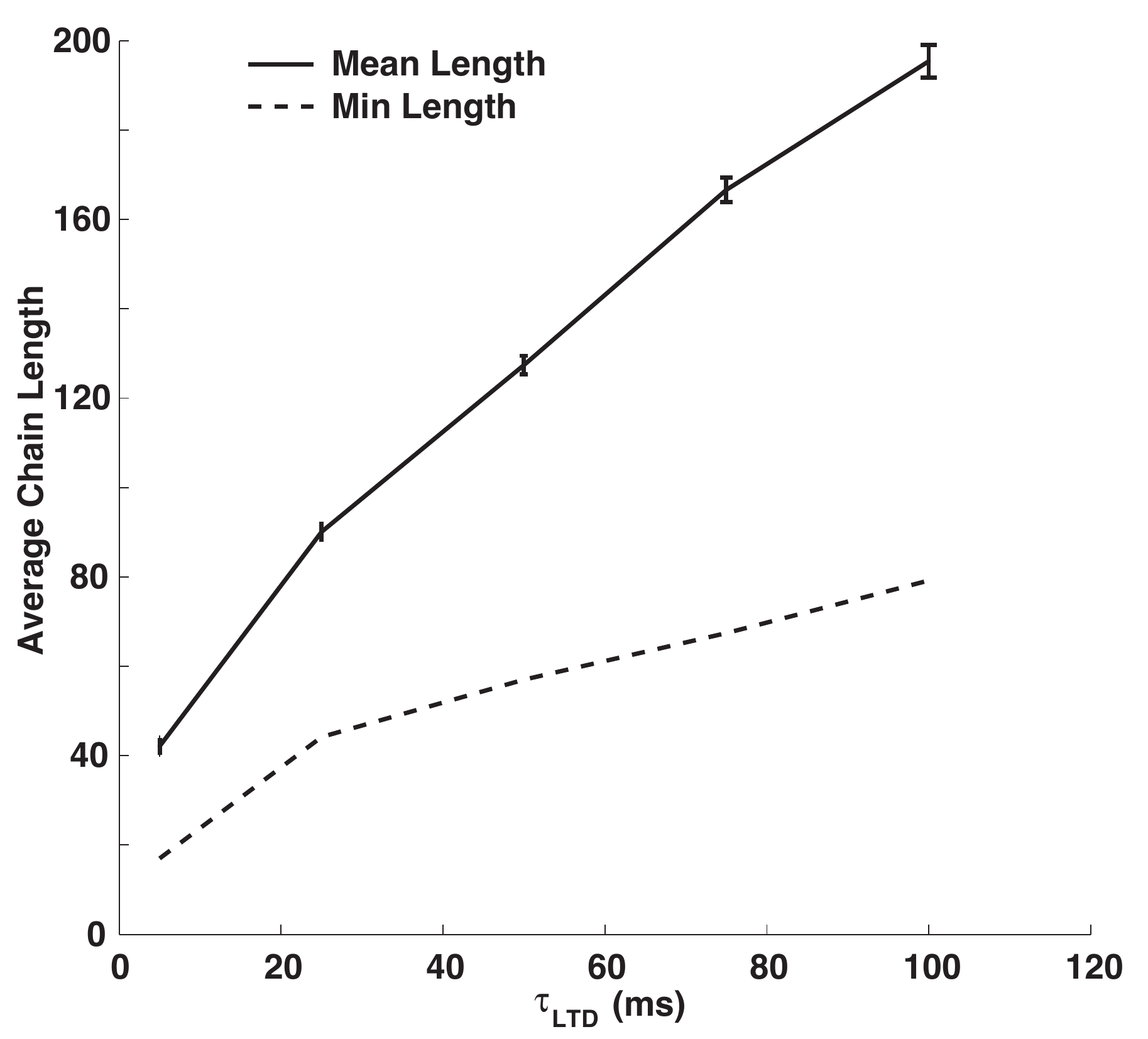}
\end{center}
\caption[LTD window result.]{}
\label{fig_LTD_shift}
\end{figure}

\begin{figure}
\begin{center}
\includegraphics[width=8cm]{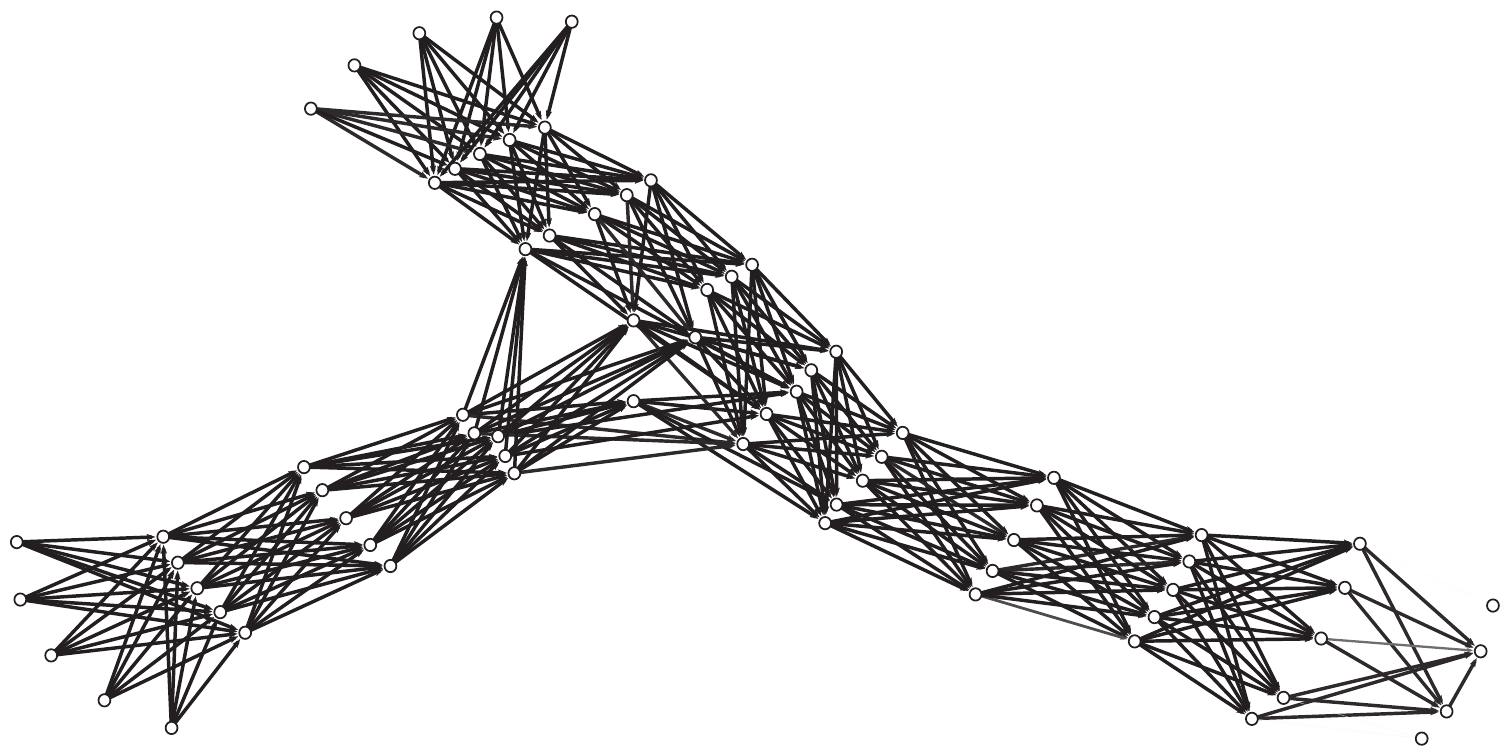}
\end{center}
\caption[Network diagram with two training sets.]{}
\label{two_training}
\end{figure}

\noindent
FIG.6.
{
{\bf (a)}
An example of re-ignition of a developing chain during the spontaneous activity period in a training trial. 
Spike raster of a network of 400 neurons are shown. 
At $0$ ms, the training neurons are stimulated, and spikes propagate down the chain until around $150$ ms. 
Spontaneous activity starts afterwards. 
The chain is re-ignited around $600$ ms, evident from the sequential spikes shown in the shaded area. 
The re-ignition starts from a random point in the chain. 
{\bf (b)} 
Spike probability for 1000 neurons is plotted against neuron label
over $10^5$ ms of spontaneous activity in a network with synaptic weights held fixed (no synaptic plasticity). 
The neurons are wired such that a synfire chain is embedded in an otherwise randomly connected network. 
Neurons labeled 1 through 130 are connected into a synfire chain,
with 1-5 forming the first group and 126-130 forming the last.
Synaptic weights of all other connections are drawn from the the synaptic weight distribution in the dynamical ground state (Fig.\ref{necessary_decay}a). 
Random spiking of neurons in the synfire chain often leads to re-ignition and propagation of spikes down the chain. 
This makes neurons at the end of the synfire chain have the highest
probability of spiking.  
The dashed line is the uniform spike probability expected in the absence of the embedded synfire chain.
{\bf (c)} Preferential targeting emerging during training is measured by
averaging the sum of active synaptic weights targeting the chain over the length of the chain.
This is plotted as a function of partial chain length over $5 \times 10^4$ training trials
for two different values of the potentiation decay. The ``enhancement'' is calculated as the sum
of weights divided by the average sum of weights onto a pool neuron in the stable  ground state.
This measurement was repeated over 10 independent instances of the network. 
Also measured and averaged (dotted lines) is the sum of weights targeting random pool neurons.
Pool neurons are more likely to target chain neurons than other pool neurons and the likelihood increases
as the chain grows.
}\\

\noindent
FIG.7.
{Chain length distributions from the lottery growth model.
{\bf (a)}
Chain length distributions are plotted for different probabilities $p_0$ from the pool neurons to the chain. 
As $p_0$ decrease, the mean  and the variance of the distribution increase. 
For $p_0=0$, there are no preferential targeting, and the mean length and variance are maximal.
{\bf (b)}
Comparison of the mean chain length as a function of the network size between the model and the simulations. 
The simulations were done for five network sizes. 
For each network size, 150 simulations were performed.
The data points are the mean chain length and the error bars indicate the standard error of the mean. 
The model prediction is plotted as the solid line. 
The parameter $p_0$ in the model was picked such that
the root-mean-square error between the predictions and the simulations at the five network sizes is minimized. 
The dotted line is the prediction for $p_0=0$, for comparison. 
}\\

\noindent
FIG.8.
{Mean chain length is offset with the LTD window size $\tau_{LTD}$.}\\

\noindent
FIG.9.
{Simulations of a 2500 neuron network with two training sets. 
Preferential targeting causes the two chains to merge.
}

\newpage 

\end{document}